\documentclass[a4paper, 11pt]{article}

\newcommand{\rsq}{\text{R}^2}
\newcommand{\ke}{\langle k_e \rangle}
\usepackage[margin=1in]{geometry} 

\usepackage{amsmath}
\usepackage{amsthm}
\usepackage{amssymb}

\usepackage[utf8]{inputenc}
\usepackage{hyperref}
\hypersetup{
	unicode,
	pdfauthor={Santosh Manicka, Manuel Marques-Pita, Luis M. Rocha},
	pdftitle={Effective connectivity determines the critical dynamics of biochemical networks},
	pdfsubject={Complex Networks and Systems Biology},
	pdfkeywords={Complex networks, criticality, controllability, generic regulation},
	pdfproducer={LaTeX},
	pdfcreator={pdflatex}
}

\usepackage[sort&compress,numbers,square]{natbib}
\theoremstyle{plain}

\theoremstyle{definition}

\usepackage{graphicx, color}
\graphicspath{{fig/}}

\usepackage{algorithm, algpseudocode} 
\usepackage{mathrsfs} 

\usepackage{lipsum}

\title{Effective connectivity determines the critical dynamics of biochemical networks}
\author{Santosh Manicka$^{1,2,+}$ \and Manuel Marques-Pita$^{1,2,3,+}${\footnote{{$^+$ These authors contributed equally to this work.}}} \and Luis M. Rocha$^{1,2}$}

\date{
	$^1$Center for Social and Biomedical Complexity, Luddy School of Informatics, Computing \& Engineering, Indiana University, Bloomington IN, USA \\ 
	$^2$Instituto Gulbenkian de Ciência, 2780-156, Oeiras, Portugal \\
	$^3$Universidade Lusófona, COPELABS. 1700-097, Lisbon, Portugal
	\\ \texttt{rocha@indiana.edu}\\[2ex]%
}

\begin{document}
	\maketitle

\begin{abstract}
Living systems operate in a critical dynamical regime---between order and chaos---where they are both resilient to perturbation, and flexible enough to evolve. 
To characterize such critical dynamics, the established \textit{structural theory} of criticality uses automata network connectivity and node bias (to be on or off) as tuning parameters.
This parsimony in the number of parameters needed sometimes leads to uncertain predictions about the dynamical regime of both random and systems biology models of biochemical regulation. 
We derive a more accurate theory of criticality by accounting for \textit{canalization}, the existence of redundancy that buffers automata response to inputs.
%
%
%
The new \textit{canalization theory} of criticality is based on a measure of \textit{effective connectivity.} 
It contributes to resolving the problem of finding precise ways to design or control network models of biochemical regulation for desired dynamical behavior. 
Our analyses reveal that effective connectivity significantly improves the prediction of critical behavior in random automata network ensembles. 
We also show that the average effective connectivity of a large battery of systems biology models is much lower than the connectivity of their original interaction structure. This suggests that canalization has been selected to dynamically reduce and homogenize the seemingly heterogeneous connectivity of biochemical networks.
		
	\end{abstract}


\section*{Introduction}

In the study of biological, social, and technological systems, network models have become important tools. \cite{barabasi2016network}  
Network model structure is defined by a graph $\mathcal{G} \equiv (X,E)$, where actors (e.g., biochemical species and environmental factors) are represented as a set of \emph{nodes}, $X$, and interactions between pairs of nodes as a set of \emph{edges}, $E$.
Several dynamical properties of networks can be inferred from their structure alone. \cite{barrat2008dynamical}
The structural properties of networks yield insights into the organization of living systems and societies. \cite{barabasi2016network, BarabSFRev3}
Yet, the rules of interaction between nodes must be considered in order to study network dynamics.
For biochemical systems, understanding the precise inference of interaction rules is a difficult task because vast amounts of data are required to estimate the kinetic parameters governing molecular concentration rates.
In response, a growing number of successful modelers have overcome the need for precise parameter estimation by relying on coarse-grained qualitative approaches to modeling interactions between nodes.
 \cite{strogatz2018nonlinear,RekAlb3,peter2012predictive, helikar2012cell, albert_othmer_2003topology}

In 1969 Kauffman introduced the simplest qualitative model of biochemical regulation and signaling, the Boolean network (BN).  \cite{Kauffman_RBN_69}
Nodes in a Kauffman network are defined as simple Boolean automata, and consequently, their interactions are described as logical state transition rules.
The state transition rule of a node, $x_i,$ incorporates $k_i$ inputs---typically the states of other nodes, or external signals.
Network dynamics ensue as the state of every node $x_i \in X$  is updated synchronously in discrete time steps.
As the dynamics unfold from an initial configuration, the network eventually settles into an \emph{attractor} configuration. 
An attractor can be a stable fixed-point---a network configuration that leads to itself in the next time step---or a sequence of configurations that repeat periodically.
Kauffman observed that hard-to-predict behaviors of real-world biochemical systems are also exhibited by these canonical networks, and they have since been widely used to model genetic regulation and signaling. \cite{OO,albert_othmer_2003topology,KineticLogic,Bornh,SchmuCrit,RekAlb2,helikar2012cell} 
State transition rules in these models are derived from molecular data and used to capture the characteristic combinatorial regulation pervasive in biochemical networks. \cite{guet2002combinatorial,macia2009cellular, davidich2008transition,RekAlb2, chaves2005robustness, alon2019introduction} 
Attractors then correspond to stable states of real systems, such as those that determine cell fate, as evidenced in a large number of models.  \cite{OO,macia2009cellular,davidich2008transition,Bornholdt_Kauffman2019, CellCollective}

Important discoveries in biology have been made using automata models even though they are built from coarse qualitative representations of biochemical entities and interactions and they use state-transition rules that often ignore the precise specification of interaction timings. \cite{RekAlb2, Bornh, yang2016compensatory, macia2009cellular}
For instance, the BN model of the yeast cell cycle reproduced the complete dynamical trajectories originating from known initial conditions to known attractor configurations. \cite{davidich2008boolean}
For another instance, a BN model of intra-cellular signal transduction in breast cancer reproduced known drug resistance mechanisms and uncovered new and effective drug interventions. \cite{zanudo2017breastCancer} 
A third, striking instance is a prescription obtained from a BN model for how to reprogram already differentiated cells. \cite{chang2011reprogramming}
While most existing models of biochemical regulation are Boolean, an increasing number of models have considered automata with more than two states. \cite{RekAlb2,zanudo2018cancer}
In addition, automata networks have been extended to include various sources of stochasticity. \cite{Bornh, CanCtrl}
BNs have become important conceptual frameworks to study a number of general principles in theoretical biology, two of which, \emph{canalization} \cite{waddington:1942,CanCtrl} and \emph{criticality}, \cite{Kauffman_RBN_69,OO,langton1986studying, langton1990computation} are at the core of the research presented here.

\subsection*{Critical dynamics in complex networks}

The notion of criticality emerged from the observation that some dynamical systems can be in a state of thermodynamic equilibrium that depends on some \emph{critical} parameter.
Tuning this parameter makes the system undergo phase transitions. 
In an \emph{ordered} phase, the system becomes insensitive to perturbations and changes in initial conditions.
Conversely, in a  \emph{chaotic} phase, dynamic trajectories within the system vary vastly as a result of small perturbations or minute differences in initial conditions. 
In the \emph{critical} phase---the one between order and chaos---the system is robust to most small perturbations, yet sensitive to some, making it flexible enough to respond differentially to a range of input signals.
In this phase, small changes in initial conditions do not lead to completely different dynamic trajectories. 
Though other notions of criticality exist, this is the focus of the research presented here.
In theory, complex networks in the critical phase can perform collective information processing, which may be a key aspect in complex life processes, such as genetic regulation. \cite{packard1988adaptation,crutchfield1988computation,langton1990computation, kauffman:1984}
%

%

\subsection*{Criticality in living systems}

Kauffman \cite{Kauffman_RBN_69} not only introduced random BNs (RBNs) to model genetic regulation, but also presented one of the first intuitions about criticality in living systems.
He suggested that biologically plausible regulatory networks that exhibit the kinds of stable dynamics seen in biology, must have, on average, \emph{low} connectivity.
Later, Kauffman elaborated on this intuition by proposing the hypothesis that biological systems operate in a \emph{critical dynamical regime}, between order and chaos, and that a critical tuning parameter is the network connectivity.
Kauffman hypothesized that each biochemical species in a given regulatory network should have two regulators on average---that is, a regular network connectivity where every state transition rule incorporates $k \approx 2$ inputs. \cite{kauffman:1984, OO}
Fifty years after the publication of the original RBN paper, Kauffman and Bornholdt revisited the main claims. \cite{Bornholdt_Kauffman2019} 
They noted that the \emph{attractor hypothesis}---network attractors correspond to cell types in genetic networks---had become an accepted fact.
%
%
Kauffman and Bornholdt examined the research on the biochemical criticality hypothesis and highlighted the following supporting evidence:
(a) the distribution of genes damaged by the spreading effects of deleting selected genes in yeast mutant has a power law distribution, which indicates criticality; \cite{Ramo2006,Serra2007} 
(b) similar, biologically-plausible initial configurations in global gene expression data obtained from macrophages follow somewhat parallel trajectories to attractors. These trajectories are neither identical, which would indicate order, nor divergent, which would indicate chaos; \cite{Nykter2008} 
(c) a large battery of sixty-seven Boolean models of real biochemical networks operate in the critical regime based on the analysis of their predicted structural and dynamical properties. \cite{daniels2018criticality}
Indeed, it is now widely accepted that biochemical complex networks are critical. \cite{Hidalgo2014,Krotov2014, EukCrit2, BallezaCrit, SchmuCrit} 
See Roli et al.\cite{roli2018criticalityreview} and Muñoz\cite{munoz2018colloquium} for recent reviews that explore further evidence of criticality in living systems.

\subsection*{The parameters and quantification of criticality} 

Several methods have been proposed for quantifying criticality in complex networks.
Early developments were grounded in physics and dynamical systems theory.
For example, Bak \cite{Bak1988} showed that regular, spatially-extended, dissipative dynamical systems can evolve to a self-organized critical state with spatial and temporal power-law scaling behavior.
%
%
%
%
Around the same period, Langton studied network dynamics using computer simulations of cellular automata (CA)---a canonical discrete idealization of spatially extended systems. \cite{langton1986studying, langton1990computation}
%
%
Langton made a very important connection between the local behavior of CA state transition rules and the collective phenomenon of criticality.
He introduced a (local) parameter, $\lambda$, to measure the proportion of state transitions in a given CA rule that do not go to a basal (quiescent) state.
CA with different values of $\lambda$ have collective behaviors that closely match distinct classes.
In Langton's account, the transition between order and chaos takes place at $\lambda \approx 0.5$ where he observed the properties of a second-order phase transition, such as in power-law distributed transients, and the maximization of average mutual information between cells.

In the same period, Derrida and collaborators made very similar connections between network parameters and collective critical behavior, but instead of looking at spatially extended cellular automata---characterized by having fixed and regular network structure---they focused on Kauffman's RBNs, where nodes have distinct state-transition functions, and networks have heterogeneous structure. \cite{DMOrig1,DMOrig2}
Indeed, Derrida and Pomeau defined what we refer to as the \emph{structural theory} (ST) of criticality for BNs. \cite{DMOrig2} 
According to this theory, if a BN has homogeneous in-degree, $k,$ and fixed bias, $p$, then the critical boundary between ordered and chaotic network dynamics is given by, 

\begin{equation}
\label{eq:currentTheory}
2kp(1-p)=1.
\end{equation}
%
%
The ST, as  defined originally, holds for fixed connectivity and fixed bias. It has since been shown that the same theory holds when connectivity is not fixed, but normally distributed around a characteristic mean value. The same applies to the bias. \cite{ALD}
While Eq.~\ref{eq:currentTheory} is theoretically well-founded, 
it is not an accurate predictor of dynamical regime, particularly if the BN dynamics are near the critical edge.
This is the case even for BNs that abide the most strict, fixed in-degree and bias assumptions.
We elaborate on this in the following sections.

While Langton determined dynamical regime using properties such as transient lengths, number of attractors, and attractor sizes in CA, Derrida and colleagues formalized a quantitative collective measure known as the \emph{Derrida parameter}, $\zeta$. \cite{DMOrig1}
This parameter is derived from the Derrida plot: a curve that shows the degree to which small perturbations in pairs of otherwise identical initial configurations diverge in their dynamical trajectories.
This divergence is measured as the average number of different node-states (Hamming distance) that separate the two initial trajectories after a number of time-steps.
The $\zeta$ parameter is the slope of the Derrida plot at the origin.
If $\zeta < 1$, the BN is classified in the ordered regime.
Conversely, if $\zeta>1$, it is classified as chaotic.
Thus a value $\zeta \approx 1$ indicates criticality.
We use $\zeta$ to determine the dynamical regime of BN ensembles in this research.
(See \S~Methods for details.)

%

\subsection*{Canalization}

\emph{Canalization} \cite{waddington:1942} is used to characterize the buffering of genetic and epigenetic perturbations that lead to the stability of phenotypic traits. \cite{siegal:2002,tusscher:2009}
Gene regulatory networks, for example, have the remarkable feature that they tend to be made of highly canalizing regulatory interactions. \cite{OO, EukCrit2, BallezaCrit,daniels2018criticality}
Canalization has been studied by characterizing redundancy in the state transition rules of automata. \cite{kauffman:1984,OO,BassSymm1,KaufStabl,CanCtrl} 
It is observed when an automaton's state transition can be determined from the known state of a subset of its inputs, which means the remaining inputs are contextually irrelevant or redundant. \cite{BassSymm1} 
Canalization influences the dynamical behavior of automata networks, contributes to their stability, \cite{CanalFracCrit, BallezaCrit, KaufStabl, NestCanDep, SenCrit} increases the likelihood of modular attractors, and makes canalizing networks more controllable by tuning external signals. \cite{CanCtrl,gates2016control} 
Yet, canalization has not previously been considered as a tuning parameter to define the structure and dynamics of critical networks or to find the most biologically plausible models of biochemical systems by targeting their stability and criticality.\cite{daniels2018criticality}
%

Previous studies of the effects of canalization on network stability and criticality have focused on the so-called \emph{strictly canalizing} state-transition rules. \cite{BassSymm1} 
Such rules always have one input that, in at least one of its possible states, is sufficient to determine the automaton's state transition.
The idea behind these studies was to build BNs with strictly canalizing state-transition rules, measure the \emph{average sensitivity} of their nodes, and quantify the propensity of an automaton to change its state as the result of perturbation to one of its inputs. \cite{SenCrit}
This node measure was then extended to quantify sensitivity at the network level.
Notably, the average \emph{network} sensitivity is equivalent to the ST  defined in Eq. \ref{eq:currentTheory} for predicting criticality. \cite{daniels2018criticality}
However, canalization is a much more frequent phenomenon when we consider, not only strict canalization, but also \emph{collective} canalization. \cite{BassSymm1}

In Boolean automata, collective canalization is observed when a subset of inputs, in some state combination, {\em jointly} determines an automaton's state transition. 
An automaton's \emph{effective connectivity}, $k_e$, introduced by authors Marques-Pita and Rocha,\cite{CanCtrl} is a measure of the expected minimal number of inputs that are necessary to determine its state transitions. 
It accounts for the existence of both strictly and collectively canalizing inputs.
If we consider how the original connectivity structure of a BN is affected by canalization, it becomes clear that it is not a useful representation of how control signals propagate in network dynamics because canalizing rules make some edges in the original structure contextually redundant.
The roles of some edges in transmitting control signals vary, however, in that some edges become completely redundant, or conversely, essential, in different dynamical trajectories and attractors.
There are many possible \emph{effective structures} with very distinct dynamical behaviors for any given in-degree network structure. \cite{gates2016control} 
This must be considered in order to understand how structure and dynamics account for control and criticality together. \cite{daniels2018criticality,gates2016control} 
Effective connectivity can easily be extended to a network measure by computing the mean in-degree of the effective structure of a given automata network.
The effective \emph{network} connectivity characterizes both the interaction structure and the canalization in one parameter.
%


We have addressed the limitations of the ST of criticality in automata networks \cite{daniels2018criticality} in Eq.~\ref{eq:currentTheory} by including the \emph{effective network connectivity} as a tuning parameter to account for canalization.
To test our main hypothesis that this parameter is a better predictor of criticality than the in-degree connectivity, $k$, we frame the prediction of dynamical stability as a binary classification problem.
We have produced a large dataset of random Boolean-network ensembles with nearly 300K distinct networks generated under the same assumptions made by ST concerning network connectivity, $k,$ and bias, $p$.
To include effective network connectivity as a tuning parameter, we have produced a catalog of state-transition rules for each $k_e$ value across the values of $k$ and $p$.
Furthermore, we have analyzed a large set of 63 models of biochemical regulation and signaling obtained from the \href{https://www.cellcollective.org/}{Cell Collective} repository. \cite{helikar2012cell,CellCollective}
For potential application areas, note that this repository includes automata models on, for example, lac operon interaction, T-cell receptor signaling, yeast cell cycle and apoptosis, cholesterol regulation, Influenza A replication, Drosophila body segmentation, lymphocyte differentiation, and cortical area development.

\section*{Results}

%
We define the effective connectivity of a given homogeneous BN
as the mean effective connectivity of its nodes, all of which have been sampled from a small interval of size $\Delta k_e=0.5$. 
The characteristic  mean value of every such interval is denoted by $\ke$.
In addition, based the principle of bias symmetry in logical rules, the compound term $p(1-p)$ is set as an independent variable that represents the bias parameter.
The dynamical regime of every BN in our ensemble data is characterized by a binary transformation of its Derrida parameter, $\zeta$, whereby
%
$\zeta > 1$ represents the chaotic regime and $\zeta \le 1$ 
%
represents the ordered or critical dynamical regimes (considered together for classification purposes since critical networks are found in the boundary between ordered and chaotic networks).
See \S~Methods for further details.


\begin{figure}
\centering
\includegraphics[width=.6\linewidth]{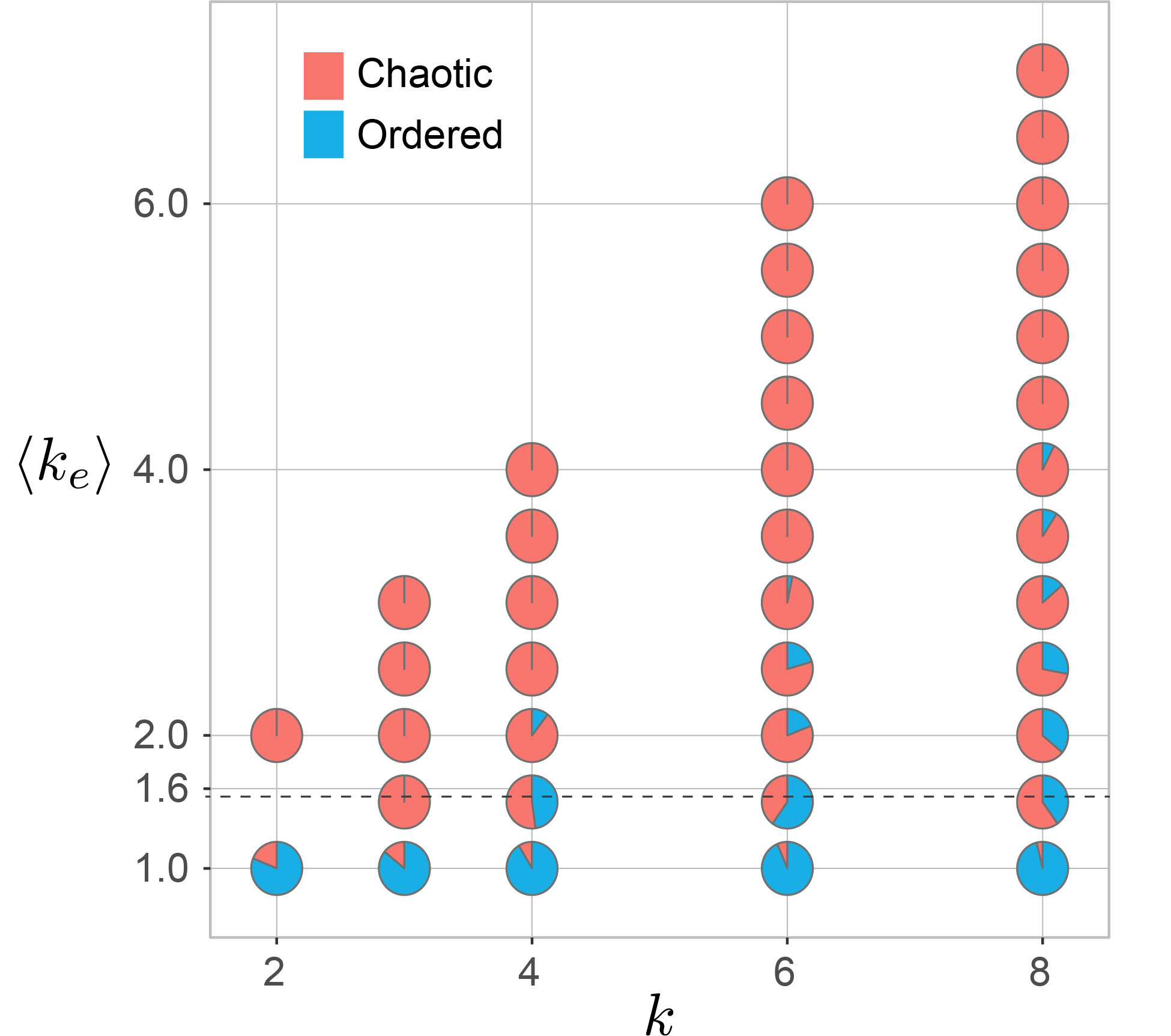}
\caption{\textbf{Dynamical regimes in the $(k,\ke)$ parameter space.} Dynamical regime (ordered or chaotic) was determined using the Derrida parameter $\zeta$ computed for the large RBN ensemble we generated (see main text). Pie charts depict the dynamical regimes of RBN aggregates for the possible $(k,\ke)$ pairs in our RBN ensembles. Blue and red areas indicate the proportions of networks with stable and chaotic dynamics, respectively. The black dashed line corresponds to the critical $\ke$, as described in the main text. The critical boundary equation, derived using binary symbolic regression in a linear model with an interaction term, is $0.1k+0.7\ke -0.1k\ke=1$. Out of the 266,400 BNs in our RBN ensembles, 224,083 (approx. $84\%)$ are classified as chaotic. 
}
\label{Fig:KKePhase}
\end{figure}

\subsection*{The canalization theory of criticality}

We search and optimize binary classifiers to predict the dynamical regime of the RBNs in our ensemble dataset using six specific model classes of increasing complexity.
For each class, there is a model instance that considers the original connectivity, $k$, and another that considers $\ke$ instead.
All other elements of a given class are kept identical in both instances; see \S~Methods for details.
Figure \ref{Fig:KKePhase} depicts the proportions of chaotic and stable BNs in our ensemble dataset for the values of $k$ and $\ke$, as well as the best criticality decision boundary we obtained when considering $k$ and $\ke$ as tuning parameters.
The majority of BNs in our ensembles are classified as chaotic $(84\%)$, based on the Derrida parameter.
Therefore, cross-validation prediction performance is best captured by measures tailored for unbalanced classification scenarios such as the \textit{Matthews Correlation Coefficient} (MCC). \cite{baldi2000assessing} 
We also show results for McFadden's $R^2$ since we are performing logistic regression, as well as the \textit{Area Under the Curve} (AUC) for ranking performance; see \S Methods for details.

\begin{figure}
\centering
\includegraphics[width=.6\linewidth]{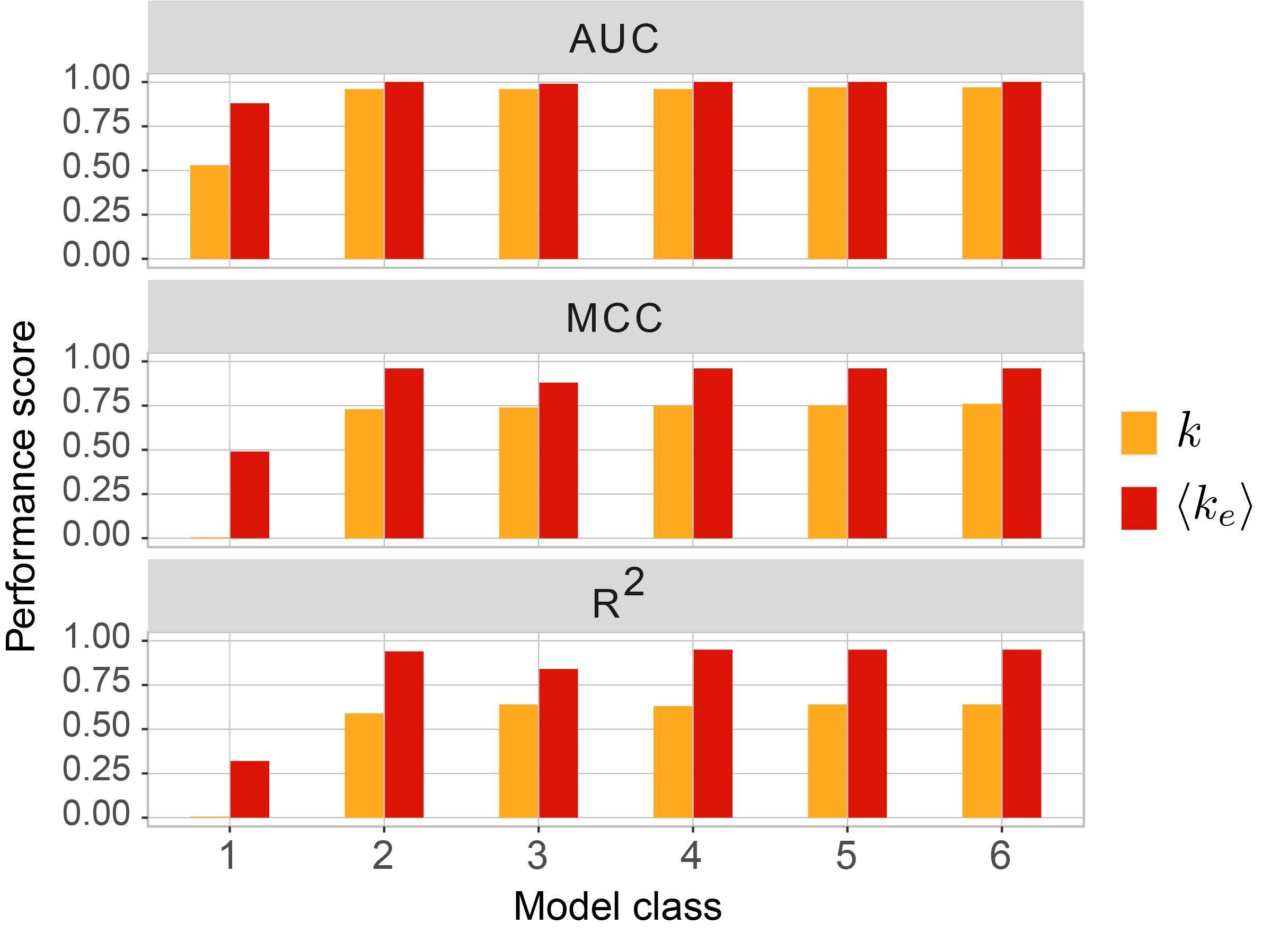}
\caption{\textbf{Performance scores for the regression models used to find the optimal critical boundary.} Each model belongs to one of six model classes---labeled in increasing order of complexity. For each model class, orange illustrates $k$ as a tuning parameter, and red, $\ke$ instead. In every class, $\ke$ is a better tuning parameter for criticality than $k$. See \S Methods for further details about classes and performance measures.}
\label{Fig:ModelPerfs}
\end{figure}

Model class (1) has  the lowest complexity, and serves to compare the predictive power of the original network connectivity, $k$, with that of the effective connectivity, $\ke$, disregarding the bias parameter.
It yields the following decision boundaries: $-0.09k=1,$ and $0.63\ke=1$. 
%
%
The corresponding critical values for the tuning parameters are $k=-11.11$, and $\ke=1.59$.
The prediction superiority of $\ke$ over $k$ is clear in this model class, since the model instance based on, $k,$ classifies every BN as chaotic, whereas the instance based on $\ke$ partitions the data into two reasonably correct dynamical regimes. 
Indeed, as shown in the first column on the left in Figure~\ref{Fig:ModelPerfs}, while MCC$(k)$ $\approx 0$, MCC$(\ke) \approx 0.49$, with similar behavior for $R^2$.
Moreover, AUC$(k) \approx 0.5$, while AUC$(\ke) \approx 0.88$.
Thus, the best classifier based solely on in-degree $k$ is equivalent to a random coin toss, while the best classifier based solely on effective connectivity $k_e$ yields reasonably good performance\footnote{To further ascertain whether $k$ and $k_e$ interact synergistically to predict criticality, 
we perform binary logistic regression considering the linear effects of $k$ and $\ke$, as well as their interaction.
The critical decision boundary of the best such classifier is $0.1k + 0.7\ke -0.1k\langle k_e \rangle = 1$.
However, the MCC $\approx 0.49$ is essentially the same as $\ke$ instance of model 1 , which demonstrates that adding $k$ does not increase the classification performance of using $k_e$ alone.
This is also clear from the best interaction model where the coefficient of $\ke$ is seven times larger than the others.}.
The study of model class (1), as well as the lack of synergy between $k$ and $k_e$, demonstrates that the original network connectivity on its own carries no information about criticality, but effective connectivity, on its own, yields a reasonable prediction of criticality. This result strongly suggests that dynamical canalization alone is an important factor in criticality.

\begin{figure*}[t]
\centering
\includegraphics[width=.99\linewidth]{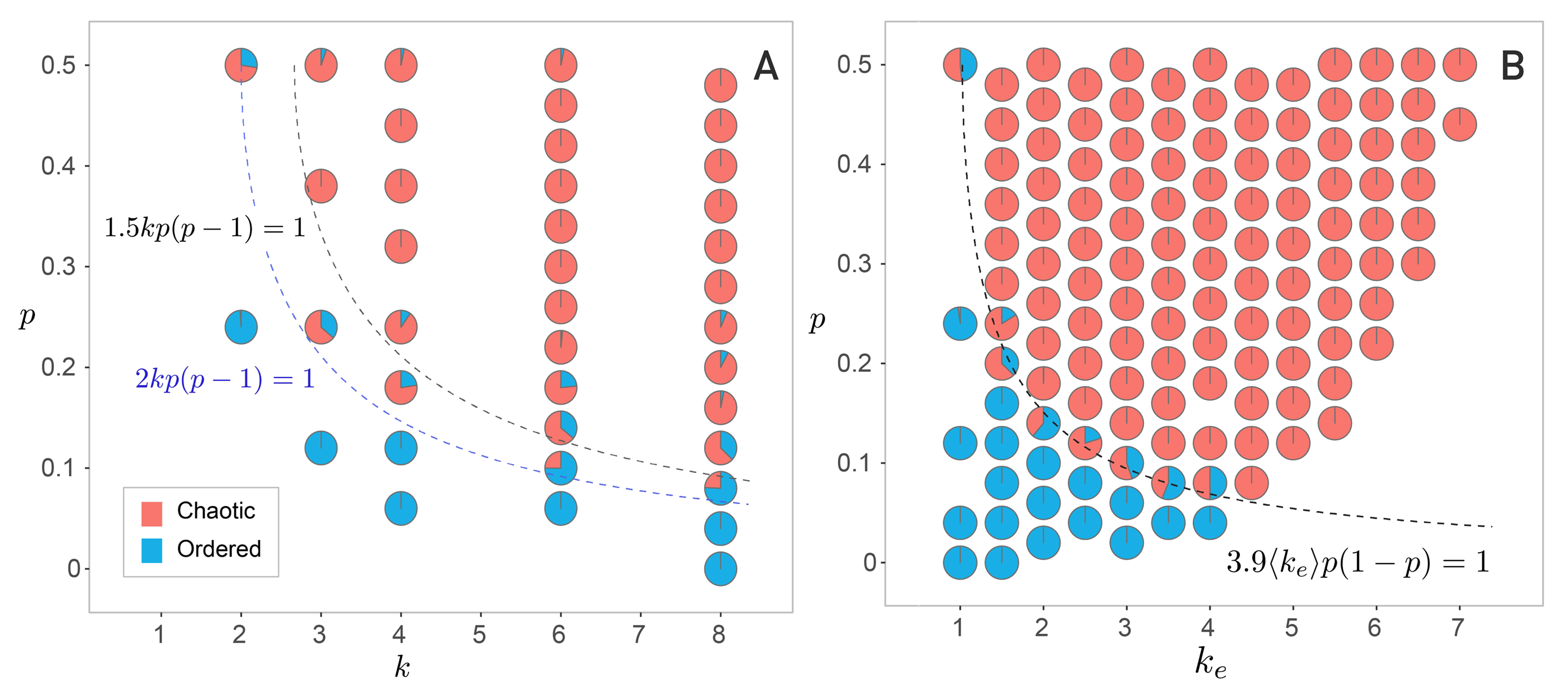}
\caption{\textbf{Dynamical regimes in the $(k, p)$ and $(\ke, p)$ parameter spaces}. Dynamical regime (ordered, or chaotic) was determined using the Derrida parameter $\zeta$ computed for the large RBN ensemble we generated (see main text). Blue areas indicate proportions of networks with ordered dynamics, and the red areas indicate the proportions that were found to be chaotic. Panel (A) depicts the $(k, p)$ space, while panel (B), the $(\ke, p)$ space. The black dashed curves represent criticality models in model class (2) described in the main text. The dashed blue curve shown in (A) corresponds to the current criticality model per the ST, shown for reference.}
\label{Fig:KPKePPhase}
\end{figure*}

Model class (2) is defined by the interaction between the term for the bias parameter $p(1-p)$ and the term for either $k$ or $\ke$ (see \S Methods).
The optimal decision boundaries obtained are: $1.49kp(1-p)=1$, and  $3.93\ke p(1-p)=1$. 
The corresponding performance metrics are shown in the second column of  Figure~\ref{Fig:ModelPerfs}.
Remarkably, all classification performance measures for model instance $\ke$ in class (2) are very high, with near-perfect MCC and $R^2$ scores, and perfect ranking performance measured by AUC, as detailed below. In contrast, the classification performance for the model instance based on $k$ in the same class is substantially, and significantly, lower (see also Figure~\ref{Fig:CVPerfs}).
To better understand the performance difference for models in class (2) consider Figure~\ref{Fig:KPKePPhase}.
First, a very crisp boundary exists between stable and chaotic dynamics in the space $(\ke, p)$; the two regimes are more neatly organized with almost no misclassifications beyond the critical boundary.
This is in sharp contrast with the less distinct boundary observed in the $(k, p)$ space around the critical boundaries, predicted by both the ST and the optimized the instance of the model based on $k$.
Indeed, in these cases substantial misclassifcations occur, whereby stable networks are observed well into the predicted chaotic regime, and vice versa, especially for the ST boundary\footnote{Notice that the ST is defined by a slightly different critical boundary (eq. \ref{eq:currentTheory}) than what we obtained by optimizing model 2 $(k)$  against random ensemble data. This is likely because the ST was derived theoretically while model 2 was derived from empirical data circumscribed to a finite $k$ range. In any case, the ST is not optimal on this range and leads to slightly worse classification performance, MCC$=0.73$ and $R^2=0.28$, than model 2 $(k)$.}.
Notice, for instance, that BNs with stable dynamics  are observed for most values of $k$ when $p=0.5$ (the most adverse value of bias for stability).
The ST predicts most of these networks to be chaotic, in Figure~\ref{Fig:KPKePPhase}(A), but in Figure~\ref{Fig:KPKePPhase}(B), in the $(\ke, p)$ space, these networks neatly cluster at $\ke=1$, right on the critical boundary.
Similar behavior occurs for all bias values.
%

The classification performance together with the observation of the arrangement of dynamical regimes around the critical boundaries in Figure~\ref{Fig:KPKePPhase} demonstrate
that using effective connectivity ($\ke$) instead of the original connectivity ($k$) in RBNs leads to a much more accurate, near-perfect prediction of the critical boundary that separates stable and chaotic dynamics, as well as a more organized characterization of both regimes.
In other words, accounting for  canalization and interaction bias at the node- or micro-level, leads to optimal prediction of macro-level dynamics. 
Indeed, model 2 $(\ke)$ shows the most accurate decision boundary for the critical boundary, with more complex models yielding no increase in classification performance.
We refer to this model as the \emph{canalization theory} (CT) of criticality in BNs.

%

\subsection*{The Canalization Theory optimizes complexity and classification performance }

\begin{figure}
\centering
\includegraphics[width=.6\linewidth]{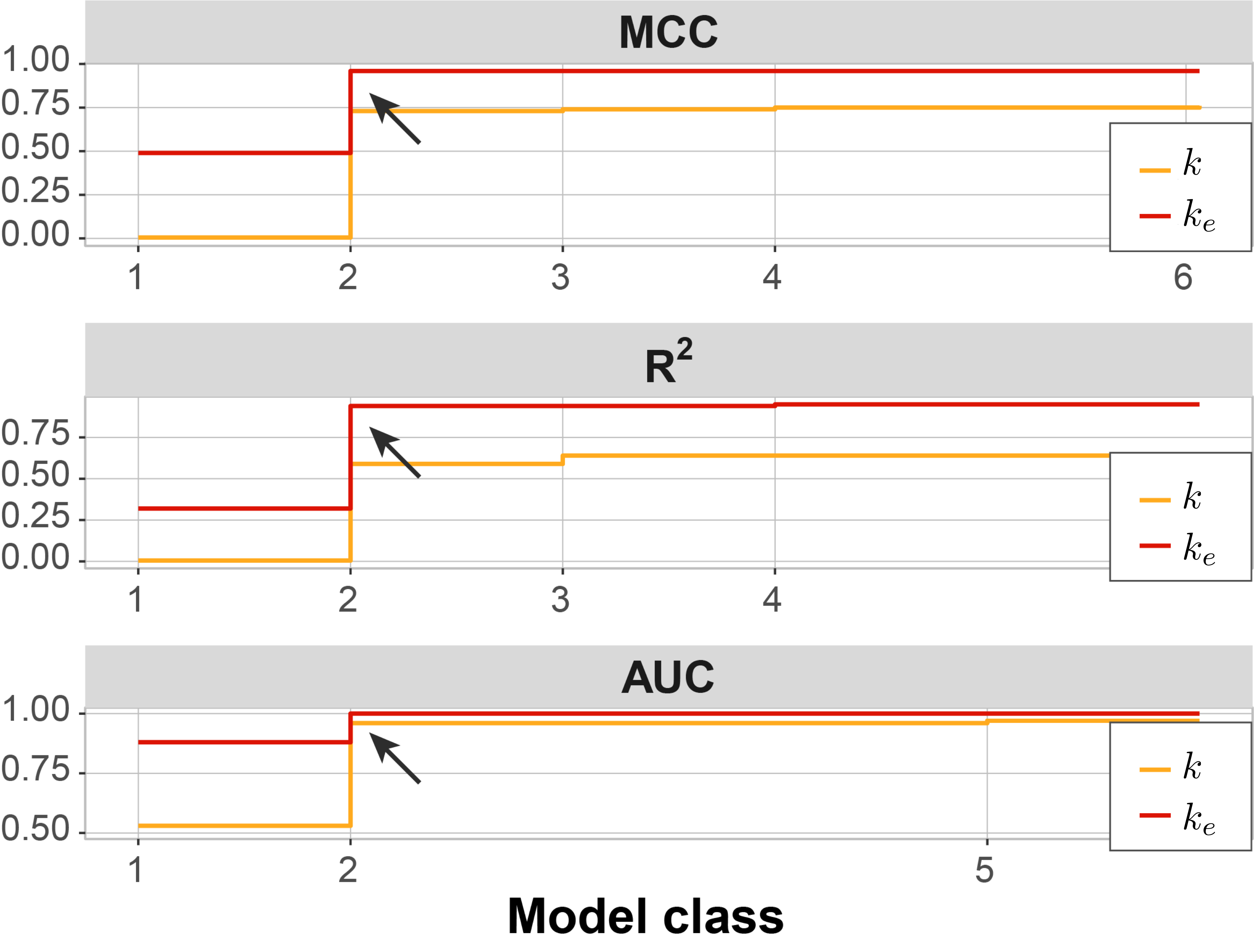}
\caption{\textbf{Pareto front analysis of model complexity vs. performance for the six model classes fit to RBN ensembles.} Models are in increasing order of complexity from class 1 to class 6. A model class is labeled on the axis only if its performance is greater than the performances of all models of lower complexity. For each model class, orange illustrates $k$ as a tuning parameter, and red, $\ke$ instead. Arrows mark the performance of the optimal model class, characterized by a substantial rise followed by very little gain afterward. Notice that for all performance measures, model class 2 with $\ke$ has the best Pareto front performance. }
\label{Fig:Pareto}
\end{figure}

We use a Pareto front method to optimize for decision boundaries that best balance the trade-off between model complexity and classification performance.
This method relies on the graphical representation shown in Figure~\ref{Fig:Pareto}, which depicts the performance of decision boundaries obtained from the different model classes considered.
A given boundary is marked with an arrow if and only if its performance is greater than that of all models of lower complexity.
In short, performance increases substantially when passing from model class (1) to (2), but not by using more complex model classes (3) and beyond.
%
%
Indeed, in model 2 $(\ke)$, the CT, achieves near-perfect classification performance with $\text{MCC}=0.96$ and $R^2=0.94$, and perfect ranking AUC $ \approx 1$. Therefore, more complex model classes could not improve much at all over such performance.
Interestingly, models based on $k$ do not show much improvement in performance beyond model complexity class (2), even though the performance of model 2 ($k$) is much smaller than that of 2 $(\ke)$ with MCC $=0.73$ and $R^2=0.58$---the former is 32\% and the latter is 64\% smaller than the respective values for the model 2 instance based on $\ke$.
In other words, even though there is much room to improve, increasing the complexity of the models based on the original connectivity, $k,$ does not yield performance gains. This implies that unless canalization is factored in, as in the model instances based on $\ke$, no increase in performance is gained over the ST.
%
%
%
%
%
%
%
%
We thus conclude that model class (2) is optimal in terms of simplicity and performance for both instances, but the instance that uses $\ke$ is considerably (and significantly as shown below) better at predicting the dynamical regime of BNs.

\subsection*{The classification performance of the CT is significantly better  than that of the ST}

\begin{figure}
\centering
\includegraphics[width=.6\linewidth]{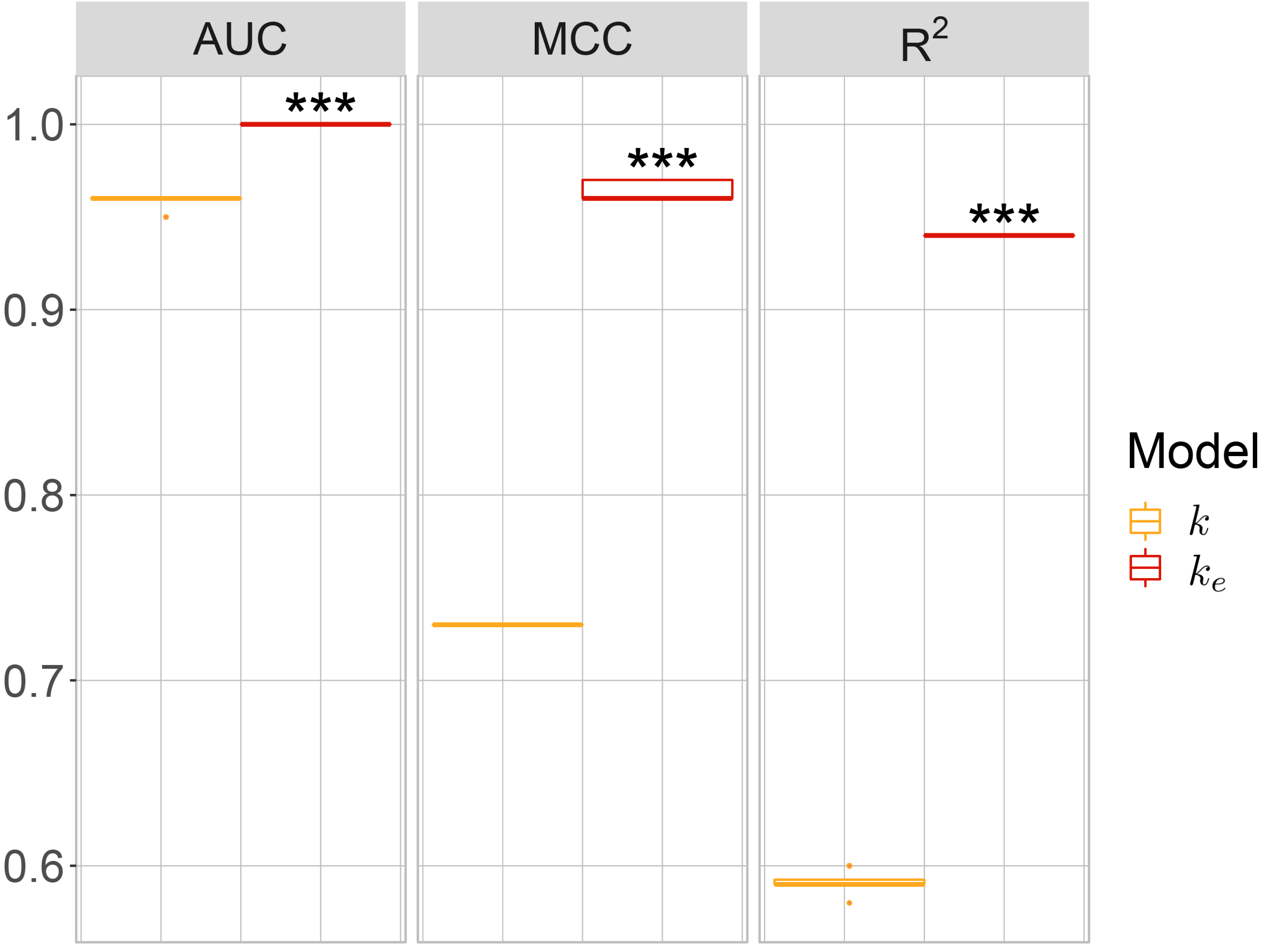}
\caption{\textbf{Classification performance of models in class 2 under nested 4-fold cross-validation.} Significant differences ($P<0.001$) are indicated with '***'. We use a one-sided paired-sample t-test to account for the alternative hypothesis that the mean score of the CT (model with $k_e$) is \emph{greater} than that of the best instance of model class 2 with $(k)$---a class that includes the ST.}
\label{Fig:CVPerfs}
\end{figure}

To estimate the statistical significance of the increased performance of the CT, as well as 
to ensure that it does not derive from over-fitting the data, we compare
both instances of models in every class under cross validation (details in \S Methods).
The statistical significance results for class (2) are shown in Figure~\ref{Fig:CVPerfs}. 
All performance measures for the CT are significantly better than for the model instance using $k$, based on paired-sample t-tests $(P < 0.001$). 
%
%
In addition, Vuong and Clarke tests and indicate similar results. 
%
%
%
These cross-validation results also demonstrate that the performance of the CT generalizes very well to out-of-sample data.
All together, this analysis supports the assertion that the CT predicts criticality in BNs significantly better than does the ST. 
See S1 in the SI appendix for details.

\subsection*{The CT via unconstrained symbolic regression}

We performed an unconstrained search using Symbolic Regression \cite{Schmidt:2009fk} alongside the constrained search reported previously in this section.
We used the symmetric effect of biases $p$ and $1-p$ on the Derrida parameter $\zeta$ to justify using only $ 0 < p \le 0.5$ rather than using the compound term $p(1-p)$. 
This type of unconstrained search works in a much larger space of model classes, so finding an optimal model that also guarantees minimal class complexity can be hard. 
Interpreting the fitting functions and coefficients can also be difficult, as stochastic searches sometimes introduce artifacts in the classifier. 
Despite these potentially limiting aspects inherent to stochastic search algorithms, we obtained a high-performance classifier that belongs to the same model class as the ST and the CT. 
The decision boundary for this classifier is the function $3.125 \langle k_e \rangle p = 1$. 
The performance of this classifier is $R^2 = 0.88$, and $\text{MCC}=0.93$, values very similar to those of the CT.
Additional information about the top classifiers produced by symbolic regression is shown in appendix S2. 
%


\subsection*{The dynamics of systems biology models is very canalized and better characterized by the CT}

\begin{figure}
\centering
\includegraphics[width=0.975\linewidth]{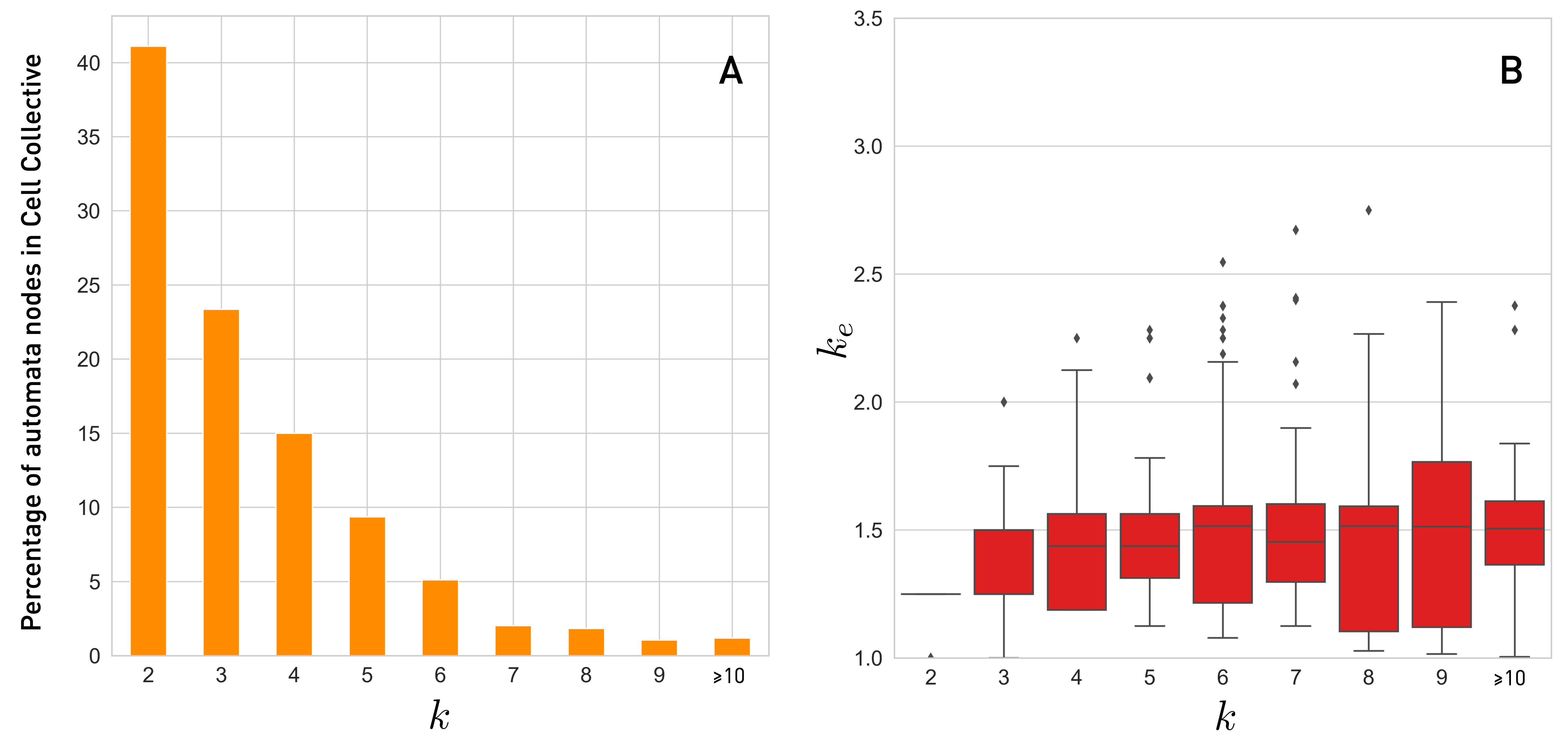}
\caption{\textbf{Characterization of $k$ and $k_e$ for the automata in the  63 Cell Collective BN models analyzed.} Depicted is the subset of automata that have two or more inputs $(52\% of the total)$, denoted in the main text by $C^*$. The median value for $k$ is Med$(C_k^*)=3$, while for $k_e$, Med$(C_{k_e}^*)=1.25$. The low median values (and low dispersion for $k_e$; see main text) indicate, not only that there is a pervasive canalization in validated BN models of biochemical systems, but also that effective connectivity `flattens' the original degree distributions. On average, knowing the state of 1.25 inputs is sufficient to determine the state transitions of these automata.  
 }
\label{Fig:CellCollKKeP}
\end{figure}

We analyze 63 Boolean models  of biochemical networks that have been experimentally validated.
These models are from the \href{https://www.cellcollective.org/}{Cell Collective} repository. \cite{CellCollective} 
We refer to the set $2,979$ automata in these models that are neither tautological nor contradictory as ${C}$.
Approximately $48\%$ of the automata in ${C}$ have one input $(k=1)$ so the connectivity of these cannot be further reduced by computing $k_e$.
Automata with  $2 \le k \le 9$ inputs account for $50\%$ of $C$, and the remaining  $2\%$ have $ 10 \le k \le 15$ inputs, see Figure~\ref{Fig:CellCollKKeP}(A). 
We excluded automata with $k=1$, since they cannot be reduced, and the very few automata with $k \ge 10$, were merged into a set ${C}^*$.
%
%
%
The in-degree distribution $(k)$ of the automata in both sets $C$ and $C^*$ is highly right-skewed with skewness $\approx 2$. 
In addition, both distributions are leptokurtic, with normalized  $\text{kurtosis} \ge 5$.
We thus report the median and interquartile range as measures of central tendency and dispersion: Med$({C}_k) = 2 $, Med$({C}_k^*) = 3 $ and IQR$({C}_k)$ = IQR$({C}_k^*) = 3-1$. 
The $k_e$ distributions for automata in ${C}$ and ${C}^*$ are also heavily right-skewed with approximately the same skewness $\approx 2$, and leptokurtic, with normalized $\text{kurtosis} \ge 6$ in both cases. 
The medians and interquartile ranges for $k_e$ are: Med$({C}_{k_e}) = 1.125 $, Med$({C}_{k_e}^*) = 1.25 $, IQR$({C}_{k_e}) = 1.25 - 1$, and IQR$({C}_{k_e}^*) = 1.43-1.25$.
Figure~\ref{Fig:CellCollKKeP} shows the $k_e$ box plots of the automata in ${C}_{k_e}^*$, one for each value of $k$.
Being so heavily leptokurtic, most of the automata in ${C}^*$ have both in-degree $k$, and effective connectivity $k_e$, very close to the respective central tendency, namely $k = 3$, and $k_e \approx 1.25$.
However, the wider dispersion for $k$ suggests that effective connectivity \emph{flattened} the original in-degree distributions of the BN models considered and shows that canalization is both very high and pervasive across different systems biology models.
See appendix S4 for additional details.

\begin{figure*}[t]
\centering
\includegraphics[width=.99\linewidth]{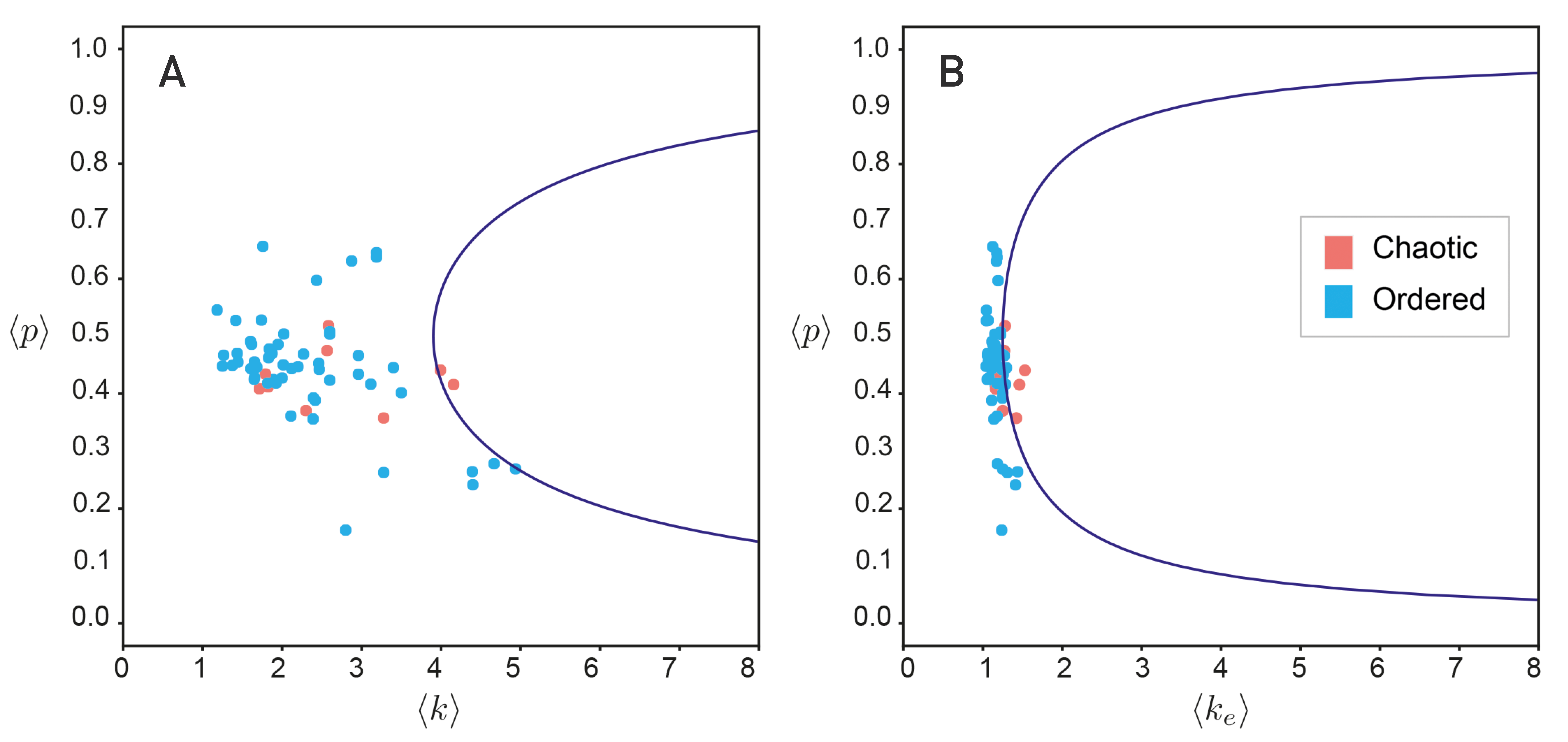}
\caption{\textbf{Predicted dynamic regimes of Cell Collective BNs by ST (panel A) and CT (panel B).} Blue dots denote stable models ($\zeta < 1$), and red dots denote chaotic models ($\zeta > 1$). The axes are labeled with the mean value of the relevant tuning parameters for each of the 63 BN models considered.
The critical boundary curves are shown in blue and have been derived by fitting class-2 models to maximize the MCC score.
%
%
}
\label{Fig:CellCollKPKePPhase}
\end{figure*}

The dynamical regime of the Cell Collective models can be inferred from their Derrida Parameter, $\zeta$, (\S Methods), which varies very little: $\text{IQR} (\zeta)= 0.976 - 0.9$ and Min/Max range $ \zeta \in [0.65, 1.15])$.
Only eleven (out of 63) models have $\zeta > 1$.
The other 52 models have $\zeta$ values slightly below $\zeta = 1$.
The low dispersion, $\zeta \approx 1$, is a strong indication that the Cell Collective models are in, or very close to, the critical regime, validating what is known about them\cite{daniels2018criticality}.  
In Figure~\ref{Fig:CellCollKPKePPhase}, we show that the near-critical status of these models is not clear in the $(\langle k \rangle,\langle p \rangle)$ space of the ST, but is quite clearly revealed in the $(\ke, \langle p \rangle)$ space of the CT.
The critical boundary curves are derived by fitting class-2 models representing the ST and the CT to maximize the MCC score.
While the networks are dispersed mostly far from the boundary curve in the ST space, they cluster very near the boundary in the CT space. Thus, the latter better characterizes the known dynamics of these models, which are mostly near critical.
Indeed, looking at the AUC ranking measure, we have $\text{AUC (ST)}=0.54$, which is only marginally better than a random toss, while $\text{AUC (CT)}=0.81$. In other words, The ranking (by distance to boundary) is far superior for the CT.
The classification performance is also superior for the CT, even though the many near-critical (and few chaotic) models make classification performance less relevant: $\text{MCC (ST)}=0.44$, $\text{MCC (CT)}=0.58$.

A caveat to this analysis of the Cell Collective models should be noted. We have developed the CT for homogeneous networks with fixed $k$ and $p$, but the Cell Collective networks are heterogeneous. Therefore, we use the mean values of these quantities in our analysis, as shown in Figure~\ref{Fig:CellCollKPKePPhase}. While the CT can be properly developed for heterogeneous networks in the future (see \S Discussion), here we derive new critical boundary curves by re-fitting both variants of model class 2 to the heterogeneous Cell Collective data.
Still, $c$ coefficients of the new curves are not very different from the optima found for the homogeneous case (Figure~\ref{Fig:KPKePPhase}): in the ($\langle k \rangle, \langle p \rangle$) space, the new $c=1.03$ (was $c=1.49$ for homogeneous case), and in the ($\langle k_e \rangle, \langle p \rangle$) space the new $c=3.2$ (for the homogeneous case, it was $c=3.93$).
The change in $c$ results in shifting the boundary curves slightly to the right in the case of the heterogeneous networks of the Cell Collective, thus increasing the area of the stable regime. This is an expected result, since we know that heterogeneous connectivity leads to more stable BN dynamics.\cite{ALD}
In summary, it is clear that including canalizing dynamics in a model of criticality yields a substantially better characterization (cf. AUC score) and prediction (cf. MCC score) of the dynamics of systems biology automata network models.

\section*{Discussion}

\subsection*{The CT is more accurate in predicting criticality than the ST and belongs to the same model class} Previous studies of criticality in automata networks have relied on the ST, which characterizes networks and their critical boundary in the $(k,p)$ space.
%
%
%
The CT introduced here includes the effects of node-level canalization and characterizes networks and their critical boundary in the $(\ke,p)$ space instead. 
In this new space, the criticality boundary leads to much more accurate predictions (Figs. \ref{Fig:ModelPerfs}, \ref{Fig:Pareto}, \& \ref{Fig:CVPerfs}), and also reveals a much more organized dynamical regime space in both random ensembles (Fig. \ref{Fig:KPKePPhase}) and systems biology models (Fig. \ref{Fig:CellCollKPKePPhase}).
Notably, the CT belongs to the same model class as the ST\footnote{We pursued both class-constrained and unconstrained regression analysis, leading to almost identical critical boundaries in the same model class}.
The Pareto-optimal model class is of the form $c \kappa p (1-p)$, where the network connectivity term $\kappa$ is the original in-degree ($k$) in the ST or the effective connectivity ($k_e$) in our new CT (See \S Methods).
The bias of state transition rules in the network is denoted $p$, and coefficient $c$ defines where the curve is positioned in the relevant parameter space (the smallest value of $\kappa$ when $p=1/2$).
Thus, in both theories, the tuning of criticality depends on interaction between the connectivity and bias parameters.
However, our work reveals that a correct measure of connectivity needs to include the influence from canalization that derives from node (state-transition) dynamics. 
Canalization at the micro-level of node dynamics defines the true connectivity of automata networks and thus ultimately their macro-level dynamical regime. 
Importantly, the prediction performance of the CT vis a vis that of the ST demonstrates that criticality depends not only on structural connectivity and bias, but also very significantly on canalizing dynamics.
Indeed, a prediction of criticality without bias (model class 1 in \S Results) shows that effective connectivity alone yields a reasonable prediction performance, but in-degree alone does not (\S Results and Figs. \ref{Fig:KKePhase}, \ref{Fig:ModelPerfs}, \& \ref{Fig:Pareto}).

%
%
%
%
%
%

%
\subsection*{Effective connectivity captures characteristic properties of dynamical regime}

In the space of $2^{2^k}$ possible logical rules for a given $k$, there are only $(2^k)-2$ distinct values of $p$ when tautologies and contradictions are ignored, and this number is halved when taking into account the principle of bias symmetry in Boolean functions.
The ST implicitly assumes that all functions of same $k$ and $p$  contribute in the same way to dynamical regime.
%
%
We demonstrate, however, that the finer characterization of the canalized logic of individual automata is necessary to accurately predict the dynamical regime of automata networks. In Figure \ref{Fig:KPKePPhase}(A), homogeneous networks of the same size whose nodes are automata with the exact same $k$ and $p$ are shown to have opposite dynamical regimes, even far from the critical boundary of the ST.
In contrast, when we transform the critical phase transition space to the finer characterization enabled by $k_e$, as in Figure \ref{Fig:KPKePPhase}(B), networks with the same $p$ and $\langle k_e \rangle$ almost always display the same dynamical regime---except very near the CT critical boundary---as demonstrated by a near-perfect MCC score (\S Results). 
Notice further that in this latter case, networks are not homogeneous in $k_e$ and are thus grouped by $\langle k_e \rangle$. Therefore, some variation in dynamical regime for the same $p$ and $\langle k_e \rangle$ is expected. Even so, such variation is only observed near the critical boundary, which demonstrates that $k_e$ (and its mean value in the BN) is very characteristic of the dynamical regime.
%
%
Finally, note that $k_e$ includes the contribution of collective canalization, while other measures of canalization such as \textit{sensitivity} do not (\S Methods). This means that the nonlinear effects of collective canalization are included and contribute to the finer characterization of criticality that the CT provides.



\subsection*{Effective structure is more homogeneous than original structure}

While we are aware that the ST has been extended to consider heterogeneous BNs---with, for example, power-law distributions \cite{ALD,FoxHill}---we have not yet considered such an extension for the CT.
One reason is that the BN models of biochemical regulation and random ensembles used here are not large enough to properly  distinguish heterogeneous degree distributions. \cite{broido2018scale}
Another important reason, as this study reveals, is that the original interaction structure of the BN is replaced by canalized dynamics that instantiates a more homogeneous effective structure with low-degree distributions.
Indeed, a consistent observation in our results is that  $\ke \ll k$ for most automata both in the random BN ensembles  and in the 63 heterogeneous Boolean models of biochemical regulation and signaling that we analyzed---$k_e = k$ only for the two parity functions for each $k$.\cite{CanCtrl} Furthermore, $k_e$ is significantly smaller in Cell Collective automata than for same size and bias random automata.\cite{gates_in_prep}
Therefore, the ubiquitous canalization (redundancy) present in automata nodes can dramatically alter the original interaction structure of a network, revealing a truer effective structure that takes canalizing dynamics into account.

It is known that such effective structure affects the dynamics and controlability of BNs.\cite{CanCtrl,gates2016control}
While effective structure can be easily computed \cite{correia2018cana} and used to uncover control pathways in biochemical regulation and signaling, \cite{gates_in_prep,CanCtrl} we do not yet know how its topology is organized across random and real-world networks.
The evidence presented here for the systems biology models in the Cell Collective indicates that the effective structure is much more homogeneous than the original interaction structures, as demonstrated by the small dispersion of $k_e$ values in comparison to the dispersion of $k$ (\S Results).
%
%
%
This suggests that very heterogeneous biological regulation and signaling networks (lognormal or asymptotic power-law degree distributions) may effectively function dynamically with more homogeneous and low-degree distributions.
An exhaustive study of the topology of effective structure is still needed to investigate this hypothesis. 
The present research, however, offers much evidence that the canalizing dynamics that defines an underlying effective structure is an important factor in determining critical dynamics in random and biochemical networks.

\subsection*{Beyond criticality: harnessing canalization in complex systems}
The theoretical development and experimental results we present provide a new theory of criticality that accounts for canalization, the CT.
Based on the same class of functions, the new theory does not increase the complexity of the current theory, but increases substantially and significantly the ability to accurately predict the dynamical regime of automata networks.
Given that automata networks are canonical examples of complex multivariate dynamical systems, the high classification accuracy of the new theory strongly suggests that canalization is a prime mechanism for tuning the dynamical regime of complex systems.
Indeed, our results with systems biology models suggest that canalization plays a fundamental role in the dynamics of biochemical regulation and signaling, which is missed by studying the structure of biochemical interactions alone.
Therefore, beyond the study of criticality, a precise characterization of canalization is likely to enable the tailoring of interventions in complex systems towards desirable dynamical behavior.\cite{CanCtrl,gates2016control}

The concept of effective connectivity underlying the CT integrates information about the structure and dynamics of multivariate interactions---in-degree connectivity and input redundancy in state transitions, respectively.
It implies that the behavior and function of complex systems is dictated by an \textit{effective structure} that is revealed only after removal of causal redundancy in the logic of how variables integrate input signals. This truer structure of interactions is a more accurate portrait of causal multivariate dynamics, which is more canalized than the original structure of interactions implies. This is why we find stable (or critical) dynamics in networks whose structure would be predicted by the current ST to be chaotic, and vice versa (see Figures \ref{Fig:KPKePPhase} \& \ref{Fig:CellCollKPKePPhase}).
In this sense, canalization is a network-level mechanism that can be tailored by evolution.
Going forward, the methodology can provide powerful analytical tools to uncover the causal pathways that determine control and resilience to interventions in various complex systems,\cite{correia2018cana} such as genetic regulation in biological development,\cite{CanCtrl} and treatment strategies in cancer and other diseases.\cite{gates_in_prep}

\section*{Methods}

\subsection*{Boolean automata definitions and notation}
\label{sec:methods}
A \emph{Boolean automaton} is a binary variable, $x \in \{0,1\}$, where state 0 is interpreted as \emph{false} (\emph{off} or \emph{unexpressed}), and state 1 as \emph{true} (\emph{on} or \emph{expressed}).
The states of $x$ are updated in discrete time-steps, $t$, according to a \emph{Boolean state transition rule} of $k$ inputs: $x^{t+1} = f\left(i_1^t, ..., i_k^t\right)$.
Therefore $f: \{0,1\}^k \rightarrow \{0,1\}$.
Such a rule can be defined by a \emph{Boolean logic formula} or by a \emph{look-up (truth) table} (LUT) with $2^{k}$ entries.
Each LUT entry of an automaton $x$, $f_{\alpha}$, is defined by (1) a specific \emph{condition}, which is a conjunction of $k$ inputs represented as a unique $k$-tuple of input-variable (Boolean) states, and (2) the automaton's \emph{next state} (transition) $x^{t+1}$, given the condition.
We denote the entire state transition rule of an automaton $x$ in its LUT representation as $F \equiv \{f_\alpha: \alpha = 1,...,2^k\}$.
\subsection*{Boolean networks}
A \emph{Boolean Network} (BN) is a graph $\mathcal{B} \equiv (X,E)$, where $X$ is a set of $n$ Boolean automata \emph{nodes} $x_i \in X, i=1,...,n$, and $E$ is a set of directed edges $e_{ji} \in E: x_i,x_j \in X$.
If $e_{ji} \in E$, then automaton $x_j$ is an input to automaton $x_i$, as computed by $F_i$.
$X_i = \{ x_j \in X : e_{ji} \in E\}$ which denotes the set of input automata of $x_i$.
Its cardinality, $k_i = |X_i|$, is the \emph{in-degree} of node $x_i$, which determines the size of its LUT, $|F_i| = 2^{k{_i}}$.
We refer to each entry of $F_i$ as $f_{i:\alpha}, \alpha = 1...2^{k{_i}}$.
At any given time $t$, $\mathcal{B}$ is in a specific \emph{configuration} of node states, $\boldsymbol{x}^t = [x_1, x_2,..., x_n]$.
We use the terms \emph{state} for individual automata $(x)$ and \emph{configuration} $(\boldsymbol{x})$ for the collection of states of the set of automata of $\mathcal{B}$, i.e., the collective network state.
Starting from an initial configuration, $\boldsymbol{x}^0$, the nodes of a BN are updated with a \emph{synchronous} or \emph{asynchronous} policy.
The \emph{dynamics} of $\mathcal{B}$ is thus defined by the temporal sequence of the $2^n$ possible configurations that ensue.
The transitions between configurations can be represented as a \emph{state transition graph}, $\textrm{STG}$, where each vertex is a configuration, and each directed edge denotes a transition from $\boldsymbol{x}^t$ to $\boldsymbol{x}^{t+1}$.
The STG of $\mathcal{B}$ thus encodes the network's entire \emph{dynamical landscape}.
Under the synchronous updating scheme (used in the studies reported in this paper) configurations that repeat, such that
$\boldsymbol{x}^{t+\mu} = \boldsymbol{x}^t$, are known as \emph{attractors}; \emph{fixed point} when $\mu=1$, and \emph{limit cycle}, with period $\mu$, when $\mu > 1$.
The disconnected subgraphs of a STG that lead to an attractor are known as \emph{basins of attraction}.
A BN $\mathcal{B}$ has a finite number of attractors, $b$, each denoted by $\mathcal{A}_i  :  i=1,...,b$.

\subsection*{Effective Connectivity}
The {\em effective connectivity} ($k_e$) tallies the expected number of inputs of an automaton $x_i$ that are \emph{minimally sufficient} to determine an its state transitions.
When a subset of such minimal inputs is in a certain state combination, the remaining inputs are effectively redundant---they can be in any state with no effect on the transition of $x_i$.
These effective inputs, or \emph{enputs} for short, can be identified using the schema redescription methodology introduced by Marques-Pita and Rocha, \cite{CanCtrl} which we illustrate next. 
The formula for the logic rule OR with two inputs can be written as $x = i_1 \lor i_2$. 
The Truth Table for this expression can be redescribed as {\em wildcard schemata} as follows: $F'_1 = \{(1,\#),(\#,1)\}$ and $F'_0 = \{(0,0)\}$, where $F'_1$ denotes the set of wildcard schemata that prescribe transitions to 1 (ON), and conversely, $F'_0$ denotes the wildcard schemata prescribing transitions to 0 (OFF), a set that contains only one schema in this case. 
The wildcard symbol `\#' in a schema denotes a redundant input state.
For example, $(1,\#)$ is interpreted as follows: given $i_1=1$, then the transition $x^{t+1}=1$ is guaranteed, regardless of the state of $i_2$. 
A closer look at $F'_1$ reveals that only one input is necessary to settle transitions to 1 (ON) in this example, and this is the case for the OR rule with any number of inputs. 
The entire set of  schemata for a given automaton can be used to determine its effective connectivity.
This requires the  computation of the average \emph{minimal} number of enputs necessary to determine its state transition.
Effective connectivity is computed from the upper bound on \emph{input redundancy}, \cite{CanCtrl} yielding a sum of the minimal number of enputs required to settle each of the possible $2^k$ state transitions specified in the automaton's LUT.
This value is then  divided by $2^k$ to obtain $k_e$.
For this computation we iterate over the entire LUT of the automaton; for each LUT entry we accumulate the number of enputs of the wildcard schema matched, with the largest number of wildcard symbols; once all LUT entries have been processed, the final accumulated sum is divided the the LUT size.
In our example $k_e=1.25$.
This is the case since \emph{three} of the \emph{four} look-up entries in the LUT have \emph{one} of the inputs in the \emph{on} state, which is sufficient to settle the transition, while one of the entries requires \emph{two} $(i_1= 0, i_2=0)$, so in this case $k_e = [(3 \times 1) + (1 \times 2)]/4$, see \cite{CanCtrl} for details. 
Note that $k_e \le k$ and that the higher the difference between $k_e$ and $k$, the more canalization there is in the automaton rule, and also, the lower the effective connectivity the automaton will have as a node in a BN.

Other measures of canalization in Boolean automata  exist and have been linked to criticality, such as \emph{average sensitivity}, \cite{SenCrit} and the more general \emph{c-sensitivity}.\cite{Kadelka}
Effective connectivity presents several advantages over these measures. First and foremost, it is designed to capture collective canalization,\cite{CanCtrl} a very common non-linear phenomenon in automata whereby a subset of inputs \textit{jointly} determine the state of an automaton, while rendering redundant the complement subset of inputs.\cite{BassSymm1}
In contrast, sensitivity independently aggregates the influence (\textit{activity}) of each individual input to an automaton. It is thus a linear measure of canalization.
This means that effective connectivity provides a more nuanced and realistic measurement of canalization that includes non-linear effects.\cite{Manicka:2017,gates_in_prep} For instance, even for automata of $k=2$, sensitivity does not discriminate between such common Boolean functions as conjunction/disjunction and proposition/negation: $s(x_1 \land x_2) = s(x_1 \lor x_2) = s(x_1) = s(\lnot x_1) = 1$. 
Effective connectivity, on the other hand, correctly accounts for the additional collective canalization that is present in the conjunction/disjunction (and other) functions: $k_e(x_1 \land x_2) = k_e(x_1 \lor x_2) = 5/4 = 1.25$, while $k_e(x_1) = k_e (\lnot x_1) = 1$.
Since non-linear, collective canalization increases with $k$,\cite{BassSymm1,gates_in_prep} the finer characterization of the phenomenon provided by effective connectivity becomes more relevant as well.
Interestingly, both sensitivity and effective connectivity can be easily computed from our schema description methodology,\cite{gates_in_prep} which is available in the CANA Python package. \cite{correia2018cana}
Finally, `$c$-sensitivity' \cite{Kadelka} extends sensitivity to subsets of $c$ inputs, but it results in a vector of $k$ values for each $c$, which is much less amenable to the regression analysis of criticality boundaries we pursue in this study than is the scalar value measured by $k_e$.

\subsection*{Generation of RBN ensembles}
Each of the ensembles of RBNs that we produced for this study is characterized by a set of tuning parameters, namely $(k,k_e,p)$. 
The network connectivity $k$ is a fixed (homogeneous) variable. 
This means that in our ensembles every node $x_i$ is connected to $k$ nodes.
The effective connectivity is the mean value in a small interval (bin), and the bias is also fixed (homogeneous).
Note that the values of these parameters are always homogeneously distributed, in alignment with the assumptions made by the ST in Eq.~\ref{eq:currentTheory}.
For a given value combination of  $(k,k_e,p)$ a single random BN is generated by choosing: (1) for each constituent node, a random set of $k$ input nodes; and (2) a random Boolean automaton with $k$ inputs, output-bias $p$, and effective connectivity in a small range $k_e \pm \epsilon$ from an existing catalog.
The reason for binning $k_e$ is that the possible values for this parameter vary significantly for each combination of $k$ and $p$, which leads to a sparse matrix of viable ensembles $(k,k_e,p)$, where viability is determined by the existence of Boolean state transition rules that satisfy specific combinations of the parameter values (see appendix S3 for further details).
Thus, without loss of information, we bin $k_e$ using a small bin size $\epsilon=0.25$ leading to $k_e$ being homogeneously distributed in regular intervals of size $\Delta k_e = 0.5$, and to a more dense matrix of viable ensembles.
Because the values of $k_e$ are binned, we refer to the $k_e$ tuning parameter as $\langle k_e \rangle$.
Producing a random Boolean automaton with a given $(k,p)$ is simple: (1) generate an all-zeroes vector of length $2^k$; (2) assign the state one (\emph{on}) to $(2^k)p$ LUT random entries in the resulting vector; and (3) assume the updated vector represents the state transitions of the automaton in the lexicographic order of input combinations. 
To control for $k_e$, we generate a catalog of Boolean automata with a large number of $(k,k_e,p)$ value combinations, from which automata with the appropriate parameter values are picked during the generation of the RBN ensembles. 
The catalogs for Boolean rules of $k={2,3,4}$ are exhaustive.
For larger $k$, automata are first obtained by random generation for a given $k$ and $p$, with their $k_e$ subsequently computed.
The number of possible automata for a given $k$ and $p$ is $\binom{2^k}{p(2^k)}$.
Thus, for $k>4$, the catalogs contain a random sample of $10^4$ Boolean rules for each $(k,p)$ if the total number possible is greater than $10^4$, and all the Boolean state transition rules otherwise. 
Additionally, to obtain automata with $k_e$ in ranges essentially inaccessible to random generation via $k$ and $p$ alone, we use a genetic algorithm. We refer the interested reader to appendix S3 for details. 
%
%
We have considered the following ranges for our tuning parameters: the number of nodes per network $N=100$, $k \in \{2,3,4,6,8\}$, $p = [0.01, 0.5]$ with $\Delta p = 1/2^k$, and $\langle k_e \rangle = [1, k]$ with $\Delta k_e = 0.5$. 
By sweeping the space of values for our ensemble parameters we have generated a total of 266.4K RBNs.

\subsection*{Computation of the Derrida parameter}
For a given BN, we compute the $\zeta$ parameter \cite{DMOrig1, DMOrig2, OO} by first generating $I=250$ random initial configurations, and producing an almost identical copy for each, where the copy differs only in the state of a small number $m$ of states that have been perturbed (flipped). We set this value to be a random integer $m \in [1,..,N/10]$. Second, allowing the BN to advance each pair of initial configurations (original and perturbed) for $t$ time steps; we set $t=1$. Third, computing the Hamming distance between the two resulting configurations. Fourth, for each value of $m$, averaging the Hamming distances obtained in the previous step and and plotting them against $m$ to produce the Derrida plot. Finally, fifth, calculating $\zeta$ as the slope of the Derrida plot at the origin. A value of $\zeta =1$ indicates criticality. A value above (below) this is interpreted as meaning the BN is in the chaotic (stable) dynamical regime.

\subsection*{Constrained search for decision boundaries}

The dataset we produce contains individual RBNs, each characterized by the independent variables $k, p$, and $k_e$, and with one dependent variable with value one (1) if $\zeta > 1$ (chaos), and zero (0) otherwise.
We perform binary logistic regression to identify the decision boundary separating dynamic regimes using a set of predefined model classes.
The general form of all models in every class is:
{\small \texttt{R=step(logistic(model))}}, where the output of the logistic function is the probability that the dependent variable has value one (chaotic regime).
The output of the step function is the predicted binary value of the dependent variable given a threshold $\tau = 0.5$.
If the output of the step function for the BN variables in a given model is greater than $\tau$ then the classifier predicts that BN to be in the chaotic regime, and critical/stable otherwise. 
%
%
Each model tested belongs to one of the following model classes, where $\kappa$ is the in-degree $k$ in the ST or the mean effective connectivity $\langle k_e \rangle$ in the CT, listed in increasing order of model complexity.
Model complexity is defined  by the number of terms and the number of predictors in each term (in that order):

{\footnotesize

\begin{enumerate} \bfseries
\item $c_1\kappa;$
\item $c_1\kappa p(1-p);$
\item $c_1\kappa + c_2p(1-p);$
\item $c_1\kappa + c_2\kappa p(1-p);$
\item $c_1\kappa p(1-p) + c_2p(1-p);$
\item $c_1\kappa + c_2\kappa p(1-p) + c_3p(1-p);$
\end{enumerate}

}
In our binary logistic regression we use the $p(1-p)$ as a single independent variable accounting for the bias, rather than just $p$ due to the principle of duality in Boolean logic. 
The coefficients derived for each {\em criticality model} are used to construct a decision surface.
For this, the resulting equations have been manipulated so that the independent variables and their coefficients are on the left-hand side and the value (1) on the right-hand side, thus facilitating comparisons with the ST.

\subsection*{Performance measures}

Mc-Fadden's $\rsq$ is a standard goodness-of-fit measure used for logistic regression models.
It is computed as one minus the ratio of the log-likelihood of the model to that of the intercept-only model. \cite{PseudoR} 
The maximum value of this pseudo $\rsq$ is 1. 
The MCC is ideal for computing classification performance in unbalanced scenarios,~\cite{baldi2000assessing} such as the one studied here, whereby there are many more instances of chaotic automata networks in the random ensembles than instances of stable network dynamics.
Computed for the classifier using model predictions and test data, it is defined as a function of the number of true positives (TP), false positives (FP), true negatives (TN) and false negatives (FN): $MCC = \frac{TP \times TN - FP \times FN}{\sqrt{(TP+FP)(TP+FN)(TN+FP)(TN+FN)}}$.~\cite{baldi2000assessing}
The MCC ranges between -1 and 1, where -1 indicates perfect opposite classification, 1 indicates perfect classification, and 0 indicates random classification. 
Here, the positive label is associated with the chaotic dynamical regime $R=1$, and the negative label with the stable (stable/critical) regime $R=0$.
The AUC is defined as a function of the true positive rate (TPR), the proportion of true positives in the total number of positive instances, false positive rate (FPR), and the proportion of false positives in the total number of negative instances, as follows: $AUC = \int_1^0 TPR(T)FPR'(T)dT$. 
The AUC ranges between 0 and 1, for perfectly incorrect and correct classification at the endpoints, respectively. A random classifier yields a value of $0.5$. It is interpreted as the probability with which the classifier ranks positive instances (label 1) higher than negative instances (label 0). \cite{hand2009measuring}

\subsection*{Cross-validation}

%
The full dataset was randomly split into 4 non-overlapping equally sized partitions ($75\%-25\%$ training and testing splits). This was repeated 4 times, thus yielding \emph{outer foldings}. 
A similar procedure was followed on each of the training splits, yielding a total of 16 training-testing pairs (see appendix S1 for further details). 
Measures of classification and regression performance (as with the full dataset) on the testing splits were collected.
The 16 sets of performance scores were averaged to produce an estimate of generalization performance score for each measure.
Between-model comparisons were made using pair-sample t-tests because the two models were evaluated on the same set of sixteen test folds. 
The paired t-tests were one-sided with the alternative hypothesis that the mean score of model 2 $(\ke)$ is greater than that of model 2 $(k)$.

\subsection*{Symbolic regression}
A supplemental study was performed using a different curve fitting method to find the critical decision surface.
We used symbolic regression (a type of unconstrained search), which is, in essence, a genetic programming algorithm. \cite{Schmidt:2009fk} 
The symmetric effect of the biases $p$ and $1-p$ on the Derrida parameter was used to prune the search space  by considering $0 <p \le 0.5$ only. 
Note that symbolic regression works in a much larger space of many function classes than the space of six model classes considered in our main methodology. 
Because of this, it can be hard to find an optimal function that is both consistent and guarantees minimal complexity.
Furthermore, the obtained classifiers and coefficients can be hard to interpret in some cases.
One of the relevant uses of this kind of method is to find different models for a given classification problem, for example, and compare them.
One of the benefits of this is to help in determining suitable function classes to describe a classification decision boundary.

Symbolic regression was performed on our dataset from different (random) seeds eight times.
We allowed for any formula in evolving populations that included basic arithmetic operators, coefficients, exponents, the sine, cosine, and logarithmic functions.
In every execution of the algorithm we consistently obtained a classifier with the same function form based on an interaction between $k_e$ and $p$ with a coefficient that varied slightly in different runs.
The ensembles were defined in the same way as in the main methods with the only difference that we used networks of size $N=48$ instead of $N=100$.
The best classifier found was the function $3.125\langle k_e \rangle p = 1$, with performance values very close to those of the CT.
See appendix S2 for further details.

\bibliography{effective-connectivity}



\section*{Acknowledgements}

MMP Acknowledges input and discussions about the original ideas with Prof. Christof Teuscher and Prof. Melanie Mitchell (Portland State University, USA), as well as funding provided by Funda\c{c}\~{a}o para a Ci\^{e}ncia e a Tecnologia (Portugal) grant PTDC/EIA-CCO/114108/2009. The authors thank Deborah Rocha for thorough line editing.
LMR was partially funded by the National Institutes of Health, National Library of Medicine Program, grant 01LM011945-01, by a Fulbright Commission fellowship, by NSF-NRT grant 1735095 ``Interdisciplinary Training in Complex Networks and Systems,'' and by Funda\c{c}\~{a}o para a Ci\^{e}ncia e a Tecnologia (Portugal) grant PTDC/EIA-CCO/114108/2009

\section*{Author contributions statement}

MMP and LMR conceived hypothesis and research rationale. SM, MMP and LMR  designed and executed the experiments, analyzed the data, and wrote the paper

\section*{Additional information}

\textbf{Competing interests} The authors do not declare any conflicts of interest. 
\end{document}